**Title: Resection cavity auto-contouring for patients with pediatric medulloblastoma using only CT information**

Soleil Hernandez, BS*[1,2], Callistus Nguyen, PhD[2], Skylar Gay, BS[1,2], Jack Duryea, BA[2], Rebecca Howell, PhD[1,2], David Fuentes, PhD[1,3], Jeannette Parkes, MBBCh[4], Hester Burger[5], Carlos Cardenas, PhD[6], Arnold C. Paulino, MD[7], Julianne Pollard-Larkin, PhD[1,2], Laurence Court, PhD[1,2]

(1) The University of Texas MD Anderson Cancer Center UTHealth Graduate School of Biomedical Sciences, Houston, TX

(2) Department of Radiation Physics, The University of Texas MD Anderson Cancer Center, Houston, TX

(3) Department of Radiation Oncology, Groote Schuur Hospital and University of Cape Town, Cape Town, South Africa

(4) Department Medical Physics, Groote Schuur Hospital and University of Cape Town, Cape Town, South Africa

(5) Department of Imaging Physics, The University of Texas MD Anderson Cancer Center, Houston, TX

(6) Department of Radiation Oncology, University of Alabama at Birmingham, Birmingham, AL

(7) Department of Radiation Oncology, The University of Texas MD Anderson Cancer Center, Houston, TX

**Abstract:**

**Purpose:** Target delineation for radiation therapy is a time-consuming and complex task. Autocontouring gross tumor volumes (GTVs) has been shown to increase efficiency. However, there is limited literature on post-operative target delineation, particularly for CT-based studies. To this end, we trained a CT-based autocontouring model to contour the post-operative GTV of pediatric patients with medulloblastoma.

**Methods:** 104 retrospective pediatric CT scans were used to train a GTV auto-contouring model. 80 patients were then preselected for contour visibility, continuity, and location to train an additional model. Each GTV was manually annotated with a visibility score based on the number of slices with a visible GTV (1 = <25%, 2 = 25%-50%, 3 = >50%-75%, and 4 = >75%-100%). Contrast and the contrast-to-noise ratio (CNR) were calculated for the GTV contour with respect to a cropped background image. Both models were tested on the original and pre-selected testing sets. The resulting surface and overlap metrics were calculated comparing the clinical and autocontoured GTVs and the corresponding clinical target volumes (CTVs).

**Results:** 80 patients were pre-selected to have a continuous GTV within the posterior fossa. Of these, 7, 41, 21, and 11 were visibly scored as 4, 3, 2, and 1, respectively. The contrast and CNR removed an additional 11 and 20 patients from the dataset, respectively. The Dice similarity coefficients (DSC) were 0.61 ± 0.29 and 0.67 ± 0.22 on the models without pre-selected training data and 0.55 ± 13.01 and 0.83 ± 0.17 on the models with pre-selected data, respectively. The DSC on the CTV expansions were 0.90 ± 0.13.

**Conclusion:** We automatically contoured continuous GTVs within the posterior fossa on scans that had contrast >±10 HU. CT-Based auto-contouring algorithms have potential to positively impact centers with limited MRI access.

## A | INTRODUCTION

Medulloblastoma is the most common pediatric malignant brain tumor, making up 25% and 50% of all pediatric brain tumors in high-income countries and low- and middle-income countries, respectively.[1] Treatment with curative intent includes total surgical resection of the primary tumor, followed by radiation therapy and chemotherapy. Residual post-operative tumor volume is correlated with the tumor stage and impacts treatment outcomes:[2] patients with more than 1.5 $cm^2$ of residual tumor are considered high- or poor-risk.[3]

In children older than 3 years, the standard radiation therapy for medulloblastoma involves irradiation of the craniospinal axis, followed by a boost to the surgical resection cavity with a margin. In the era of conformal radiation therapy and more advanced imaging, the treatment boost volume has evolved from treatment of the entire posterior fossa to an expansion around the resection cavity itself.[4] In pediatric medulloblastoma, the GTV is defined as the post-surgical cavity, including the tumor-brain interface, prior to resection. The CTV is defined as the tumor bed plus any residual tumor, with a 0.5- to 1.5-cm margin, and the planning target volume is a 3 to 5 mm expansion of the CTV according to institutional guidelines. The definition of the target volumes depends largely on the visibility of the tumor and tumor bed observed on pre- and post-operative imaging studies.[1,5]

Target delineation for radiation therapy treatment planning is a complex and time-consuming task, requiring expertise to translate surgical notes, pathology, and imaging studies into a 3D treatment volume.[6] For this reason, high inter-physician and inter-institution variation in target delineation has been consistently reported in the literature for many adult disease sites.[7–10] Similar results are emerging for pediatric disease sites.[6,11,12] Dietzsch et. al. developed and tested a pretreatment radiotherapy quality control program on 69 pediatric craniospinal irradiation treatment plans. They reported that 49.3% of the plans evaluated were flagged due to incorrect target volume delineation.[13] To this end, autocontouring is being pursued to expedite the target delineation process and potentially decrease inter-physician variability. There is extensive literature demonstrating varying success in automatically delineating solid GTV volumes using MRI information or multi-modality imaging information.[14–18] There is limited literature on automatically delineating post-operative GTV volumes. Previous post-operative autocontouring studies relied on multiple MRI sequences for training and achieved Dice Similarity Coefficients (DSCs) ranging from 0.75 to 0.89.[19–21] The literature on automatically delineating post-operative GTV volumes using CT scans alone is even more limited. Bi et al explored a deep learning–assisted contouring process to semi-automatically generate post-operative CTVs for non-small cell lung cancer. They reported a DSC of 0.75, a decrease in time spent contouring (33%), and a decrease in inter-physician variability.[22]

CT-Based autocontouring models have the potential to positively impact resource-constrained centers where MRI access may be limited. In an online survey conducted by Parkes et al., 93% and 82% of surveyed centers in 47 countries reported having access to CT and MRI, respectively. After only considering responses from African centers, the reported MRI access decreased to 77%.[1] The cost of purchasing equipment is high, and long-term maintenance makes up a significant portion of the overall cost. Machine maintenance can be difficult for resource-constrained centers because repairs are often carried out by companies covering a large geographical area.[23] Ekpo et al. summarized the challenges of maintaining medical imaging equipment in Nigeria. Of 61 imaging devices installed across nine hospitals, 16% were nonfunctional at the time of the survey. In addition, survey participants reported that 81% of minor faults resulted in up to 72 hours of downtime.[24] This downtime can be costly during the radiation therapy planning process. Even in hospitals with access to MRI, immediate post-operative MRI for children in resource-constrained centers can be problematic, since patients must be transported from ICU, and many require anesthetic for the scan. Post-operative MRI performed after 48-72 hours is unreliable for distinguishing between blood products and residual tumor. [25,26]

In this study, to explore automated target delineation using only CT information, we trained an autocontouring model to contour the post-operative GTV of pediatric patients treated for medulloblastoma. Further, we investigated the impact of pre-processing training and testing data to optimize model performance. Automating target delineation has the potential to expedite clinical workflows. Moreover, training a model that only relies on CT information has the potential to positively impact resource-constrained centers where MRI access is limited or unavailable.

## B | METHODS

In this study, we experimentally optimized a deep-learning model to contour resection cavities in patients with pediatric medulloblastoma using only CT information. Using clinical data, we trained multiple deep learning models to quantitatively assess the impact of pre-selecting the training and testing datasets based on contour visibility.

### B.1 | Data curation

A data set of 104 CT scans from patients treated for pediatric medulloblastoma was curated for this study. Retrospective patient data used in the testing and development of the autocontouring approach were collected following an institutional review board–approved protocol at our institution. The median (range) number of slices, slice thickness, and tube voltage peak were 347 (133-523), 2.5 (1.25-2.5) mm, and 120 (80-120) kVp, respectively. The patients in the data set had a median age of 7 years (range, 1.5-19 years) and a male-to-female ratio of 2:1. The age and sex distribution of our data set is comparable to that reported in the literature for pediatric medulloblastoma.[27]

### B.2 | Baseline model

To automatically generate the pediatric GTV auto-contouring baseline model, we divided the data set of 104 pediatric patients into training and testing sets (82 [80%] and 22 [20%], respectively) for a nn-UNet model.[28] This architecture was selected for the experiment because it has been found to be effective for limited and heterogeneous data sets.[28] One advantage of the nn-UNet model is that it generates a data signature to optimize the training hyperparameters for the data set, making the training process less sensitive to heterogeneities in the data (e.g., patient positioning, image scanning protocols, and anatomy variation with age). Using the optimized hyperparameters, a 3D full-resolution nn-UNet model was trained with five-fold cross validation to further maximize the limited data set. The performance of the auto-contouring tool was quantified with surface and overlap metrics (Dice similarity coefficient [DSC], Hausdorff distance [HD], and mean surface distance [MSD], respectively).

**B.3 Pre-selecting data**

In the second experiment, we determined whether removing GTV contours with poor contour visibility from the training or testing datasets would improve the overall model performance. First, the dataset from 2.1 was further curated to only contain patients that had a continuous GTV contour located within the posterior fossa. Then, contour visibility calculations were applied to the curated training and testing datasets (54 and 15 patients, respectively). To qualitatively select contours with higher contour visibility data, we first manually annotated each patient with an overall resection cavity visibility score. We used a 4-point scoring scale where 1 = < 25% of slices of resection cavity are visible, 2 = 25%-50% of resection cavity slices are visible, 3 = 50%-75% of resection cavity slices are visible, and 4 = >75%-100% of resection cavity slices are clearly visible. All cases were viewed and annotated in Raystation 11B[29] using a default brain window [L:35, W:100] to ensure consistent scoring across all patients.

[Figure 1]

To filter the data quantitively, the contrast and CNR ratio were calculated between the resection cavity and surrounding normal brain tissue and compared to the manually assigned visualization scores to understand which metric was the best representation of contour visibility.

The filtration workflow is outlined in Figure 1. To calculate contrast and CNR, the image was cropped using a 3D bounding box derived from a 10-pixel x 10-pixel x 2-slice 3D expansion of the GTV contour. High-attenuating areas such as the bone were assigned as NaN values so that they would not skew the mean and standard deviation calculations. The mean and standard deviation of the images were calculated within the GTV contour and the surrounding normal brain tissue using a NaN mean and standard deviation. Contrast and CNR were calculated according to Equation 1 and Equation 2, respectively, where $\bar{x}_{ROI}$ is the mean intensity value within the GTV contour, $\bar{x}_{bkg}$ is the mean intensity value within the surrounding brain tissue, and $\sigma_{bkg}$ is the standard deviation of the intensity values within the surrounding brain tissue.

$$Contrast = \bar{x}_{ROI} - \bar{x}_bkg \qquad \text{(Equation 1)}$$

$$CNR = \frac{Contrast}{Noise} = \frac{(\mu\bar{x}_{ROI} - \bar{x}_{bkg})}{\sigma_{bkg}} \qquad \text{(Equation 2)}$$

The contrast and CNR were quantified for all patients and compared to the assigned contour visibility scores to determine which metric would be a better indication of resection cavity visibility. An additional consideration was how selective each metric was, i.e., how the training and testing data could be filtered without decreasing a significant number of patients available for training and testing the model.

The selected mode of filtration was analyzed to determine a threshold of contour visibility-based pre-selection to apply to the training and testing images. The data were pre-selected based on the contour visibility criteria and used to train an nn-UNet model with five-fold cross validation. The performance of the model was quantified using surface and overlap metrics.

**B.4 | Comparing model performance**

To understand the impact of applying filtration to the training and testing datasets, we performed four experiments (Figure 2). Two autocontouring models were trained and tested on two datasets. Model 1 was originally trained on the entire dataset and model 2 was trained on a sub-set of model 1's training data, pre-selected for contour visibility, contour continuity and contour location. In experiment 1, model 1 was used to run predictions on the model 1 original testing set. Neither the training nor testing data was pre-selected based on the specified contour criteria. In experiment 2, model 2 was used to run predictions on the model 2 testing set that was pre-selected to remove contours that had poor visibility, were discontinuous or were located outside of the posterior fossa. Both the training and testing datasets were filtered using the same criteria. In experiment 3, model 1 was used to run predictions on the model 2 test set. In this scenario, the training data were untouched, and the testing data were pre-selected. In experiment 4, model 2 was used to run predictions on the model 1 testing set. In this scenario, the training data were pre-selected, and the testing dataset was not. The four scenarios were compared and analyzed using DSC, HD, MSD. The two models were further compared using an independent t-test ($p<0.05$ as statistically significant) for each test set.

[Figure 2]

**B.5 | Comparing top-performing GTV autocontours to clinical contours**

The top-performing GTV autocontouring model was determined on the basis of surface and overlap metrics. The GTV autocontours of the model's test patients were imported into the treatment planning system (Raystation 11B).[30] CTVs were created for both the autocontoured and clinical GTV. The CTV was defined as a 1.5-cm anatomic expansion of each GTV contour.[4] The CTVs were post-processed to be confined to the brain and not include the brainstem (if possible). For GTV contours immediately adjacent to the brainstem, the 1.5 cm expansion was reduced to 0.5 cm in the direction of the brainstem. To compare the resulting CTV contours, the surface and overlap metrics were quantified (DSC, HD, MSD, precision, and recall).

## C | RESULTS

### C.1 | Pre-selecting data for contour visibility

Of the 104 patients curated for the study, 80 were pre-selected to have a continuous GTV volume located within the posterior fossa and assigned a visibility score. Of these GTV contours scored for contour visibility, 7, 41, 21, and 11 patients were scored as 4, 3, 2, and 1, respectively. Figure 3 shows examples of patients who were scored as 1, 2, 3, and 4. We experimented with various levels of cropping for the contrast and CNR calculations. We calculated both parameters using cropping dimensions of 5, 10, and 20 pixels in the X and Y direction and two slices in the z direction. We found that the 10x10x2 cropping gave the most consistent contrast and CNR calculations across our dataset. Further, the cropping window provided enough surrounding brain tissue without introducing too much additional anatomy, like the skull and sinus cavities.

[Figure 3]

After calculating the contrast (Figure 4a) and CNR (Figure 4b) and plotting them against the visibility scores, we determined the appropriate threshold for contrast to be any value outside of [-10 HU, 10 HU] and CNR value outside of [-0.5, 0.1]. Figure 4c summaries the number of patients who were removed from the dataset based on each quantitative visibility metric. The contrast threshold maintained all patients with a visibility score ≤ 3 and removed 11 patients from the dataset. The contrast threshold removed 29% and 45% of visibility scores of 2 and 1, respectively. The CNR threshold was stricter, removing a total of 20 patients from the dataset. The threshold maintained all patients with a visibility score of 4, removed 7% of patients with a visibility score of 3, 43% of patients with a score of 2, and 73% of patients with a score of 1. After calculating both the contrast and CNR for each of the patient images, we found that contrast was the optimal metric, removing low-contrast contours without sacrificing too much training data.

[Figure 4]

### C.2 | Comparing model performance

The results of the four experiments were compared using DSC, HD, and MSD (Figure 5). The DSC achieved on the models trained on pre-selected training data were 0.61 ± 0.29 on the original test set and 0.67 ± 0.22 on the pre-selected test set. The DSCs achieved on the models with the full training data were 0.55 ± 13.01 on the full test set and 0.83 ± 0.16 on the pre-selected test set. The difference between the performance of the original model and the pre-selected model on the pre-selected testing set was statistically significant (p=0.02). In summary, both models performed better on the pre-selected test set than on the original test set. However, the top-performing model was the one that was trained on all data and tested on the pre-selected data.

[Figure 5]

### C.3 | Comparing top-performing GTV autocontours to clinical contours

The same CTV expansion was applied to both the clinical GTV and autocontoured GTV from the top-performing segmentation model. The resulting comparison metrics are summarized in Figure 6. The DSC, HD [mm], and MSD [mm] achieved on the GTV/CTV were 0.83 ± 0.16/0.90 ± 0.13, 8.95 ± 6.85/9.10 ± 7.00, and 1.16 ± 1.5/1.14 ± 2.00, respectively (Figure 6a). The average precision and recall for the GTV/CTV were 0.81/0.89 and 0.89/0.99, respectively. The percentages of GTV and CTV contours with a DSC > 0.90 were 33% and 73%, respectively. Figure 6b gives two examples of the DSC from the GTV contours (green) and resulting CTV contours (purple).

[Figure 6]

**D | DISCUSSION**

In summary, we trained a GTV autocontouring model that relies only on CT data. We assessed the impact of applying various visibility-based pre-selection techniques to the training and testing datasets. We trained two models, one with the entire dataset and one in which the training data was preselected to include high visibility contours that were continuous and located within the posterior fossa. We then tested each model on the original test set and a pre-selected test set. The top-performing model was the one that was trained using all data and tested on pre-selected data. The model achieved a mean DSC of 0.83 and showed the least spread in DSC, HD, and MSD across the testing dataset.

Ultimately, we elected to use contrast as the visibility threshold metric because it provided the best compromise between higher visibility data selection and the resulting dataset size. The CNR pre-selection threshold eliminated nearly 20 patients from our dataset, while contrast eliminated 10. The threshold of the contrast model was decided experimentally by assessing the relationship between qualitative visibility scores and the calculated contrast. We found that contrast values that were less than -10 HU and greater than +10 HU corresponded well with the assigned visibility score.

After comparing the performance of both models on two datasets, we found that the top-performing experiment was that in which the training data were left untouched and the testing data was pre-selected based on selected contour visibility criteria. In both testing scenarios, we found that the autocontours that achieved a low DSC score were a result of patients with GTV contours that were within the posterior fossa but not centrally located. Despite having high visibility scores and contrast, both autocontouring models struggled to contour the GTV in both the original and pre-selected test sets. Our CTV expansion results suggest that applying the same expansion to the clinical and autocontouring GTV volumes results in higher overall DSC, precision, and recall between the resulting CTV volumes. The extent of GTV contouring differences was minimized after expansion.

Automatic target delineation has the potential to expedite clinical workflows. Radiation therapy for pediatric medulloblastoma is performed in two parts. First, the entire craniospinal axis is irradiated, followed by a boost to the resection cavity with a margin for sub-clinical tumor. Consequently, treatment planning consists of delineating the cranio-spinal axis and the normal tissues, delineating the resection cavity, and generating multiple treatment plans to treat each volume. Recently, our group automated the normal tissue segmentation and plan generation process for pediatric craniospinal irradiation.[31] We plan to expand our methodology to include multi-modality-based resection cavity contouring and boost planning. While GTV contouring may be a fraction of the overall treatment planning process, integrating contour automation with our previous methods has the potential to significantly reduce treatment planning time, granting more time for other clinical tasks. Moreover, autocontouring using CT scans alone, has the potential to benefit centers with limited access to MRI.

The medical literature on automated post-resection GTV contouring using only CT information is limited. Bi et al explored a method to semi-automatically generate post-operative CTVs for non-small cell lung cancer on 19 CT scans. The physicians were asked to edit DLAC contours and compare them to manual contours. Using the DLAC method, the DSC overlap of the contours improved from 0.72 to 0.75, the contouring time decreased by 33%, and the inter-physician variability decreased.[22]

Men et al. used an encoder-decoder framework and 50 CTs with contrast to automatically contour stage 1 or 2 nasopharyngeal tumors (GTVs) and achieved a DSC of 0.81 on the primary tumor and 0.62 on the involved lymph nodes.[32] Mei et al. used an ensemble of U-Net models with spatial attention to automatically contour nasopharynx GTVs using 50 CT scans for training and testing and achieved an average DSC of 0.65.[33] We cannot directly compare our results to these experiments as our study did not include the same disease site and the latter studies reported outcomes in solid GTV volumes rather than resection cavities; however, our best model (trained on all data and tested on higher visibility data) achieved a DSC of 0.83 ± 0.16, which exceeds what has been reported in the limited GTV autosegmentation literature.

Like other studies, the success of our model was limited by the consistency of the target delineation in our training data. Variation in target volumes was due to varying deformation of the surrounding normal tissues following surgery and inter-physician variability. Inter-physician variability results from varying training experiences, unique contouring preferences, differing incorporations of clinical knowledge, and patient-specific tradeoffs between tumor control and toxicity.[34] Coles et al. highlighted inter-clinical variation in pediatric medulloblastoma target delineation after discovering ambiguities in the process at an educational meeting.[11] In our study, we used retrospective, clinical, pediatric data to autocontour the GTV volumes. Consequently, the number of patients used in our training and testing datasets was limited. Finally, all training and testing data were provided by a single institution. To this end, the model could be improved by incorporating external datasets.

## E | CONCLUSION

In conclusion, we were able to automatically contour continuous resection cavities located within the posterior fossa for patients with medulloblastoma who had less than -10 HU or greater than +10 HU of contrast calculated for the GTV with respect to the background image (majority of patients). Our results align with what has been reported for CT-based GTV autosegmentation and adds to the limited literature on the topic. The fact that the model only uses CT data could be of interest to resource-constrained centers that have limited access to MRI.

## ACKNOWLEDGMENTS

SH is supported by a Cancer Prevention and Research Institute of Texas (CPRIT) Training Award (RP210028) and the Dr. John J. Kopchick and Mrs. Charlene Kopchick Fellowship. Editorial support was provided by Ann M. Sutton, Scientific Editor, of the Research Medical Library at MD Anderson. We acknowledge the support of the High-Performance Computing for Research facility at MD Anderson for providing computational resources that have contributed to the research results reported in this article.

**CONFLICTS OF INTEREST**

Hester Burger is currently employed by Varian Medical Affairs, with a sessional lecturing position at the University of Cape Town.

## REFERENCES

1. Parkes J, Hendricks M, Ssenyonga P, et al. SIOP PODC adapted treatment recommendations for standard-risk medulloblastoma in low and middle income settings: Standard-Risk Medulloblastoma in LMIC. *Pediatr Blood Cancer*. 2015;62(4):553-564. doi:10.1002/pbc.25313

2. Harisiadis L, Chang CH. Medulloblastoma in children: A correlation between staging and results of treatment. *Int J Radiat Oncol*. 1977;2(9):833-841. doi:10.1016/0360-3016(77)90181-X

3. Zeltzer PM, Boyett JM, Finlay JL, et al. Metastasis Stage, Adjuvant Treatment, and Residual Tumor Are Prognostic Factors for Medulloblastoma in Children: Conclusions From the Children's Cancer Group 921 Randomized Phase III Study. *J Clin Oncol*. 1999;17(3):832-832. doi:10.1200/JCO.1999.17.3.832

4. Michalski JM, Janss AJ, Vezina LG, et al. Children's Oncology Group Phase III Trial of Reduced-Dose and Reduced-Volume Radiotherapy With Chemotherapy for Newly Diagnosed Average-Risk Medulloblastoma. *J Clin Oncol Off J Am Soc Clin Oncol*. 2021;39(24):2685-2697. doi:10.1200/JCO.20.02730

5. Merchant TE, Happersett L, Finlay JL, Leibel SA. Preliminary results of conformal radiation therapy for medulloblastoma. *Neuro-Oncol*. 1999;1(3):177-187.

6. Kristensen I, Agrup M, Bergström P, et al. Assessment of volume segmentation in radiotherapy of adolescents; a treatment planning study by the Swedish Workgroup for Paediatric Radiotherapy. *Acta Oncol*. 2014;53(1):126-133. doi:10.3109/0284186X.2013.782104

7. Li XA, Tai A, Arthur DW, et al. Variability of Target and Normal Structure Delineation for Breast Cancer Radiotherapy: An RTOG Multi-Institutional and Multiobserver Study. *Int J Radiat Oncol*. 2009;73(3):944-951. doi:10.1016/j.ijrobp.2008.10.034

8. Matzinger O, Poortmans P, Giraud JY, et al. Quality assurance in the 22991 EORTC ROG trial in localized prostate cancer: dummy run and individual case review. *Radiother Oncol J Eur Soc Ther Radiol Oncol*. 2009;90(3):285-290. doi:10.1016/j.radonc.2008.10.022

9. Yamamoto M, Nagata Y, Okajima K, et al. Differences in target outline delineation from CT scans of brain tumours using different methods and different observers. *Radiother Oncol J Eur Soc Ther Radiol Oncol*. 1999;50(2):151-156. doi:10.1016/s0167-8140(99)00015-8

10. Weiss E, Richter S, Krauss T, et al. Conformal radiotherapy planning of cervix carcinoma: differences in the delineation of the clinical target volume: A comparison between gynaecologic and radiation oncologists. *Radiother Oncol*. 2003;67(1):87-95. doi:10.1016/S0167-8140(02)00373-0

11. Coles CE, Hoole ACF, Harden SV, et al. Quantitative assessment of inter-clinician variability of target volume delineation for medulloblastoma: quality assurance for the SIOP PNET 4 trial protocol. *Radiother Oncol J Eur Soc Ther Radiol Oncol*. 2003;69(2):189-194. doi:10.1016/j.radonc.2003.09.009

12. Padovani L, Huchet A, Claude L, et al. Inter-clinician variability in making dosimetric decisions in pediatric treatment: a balance between efficacy and late effects. *Radiother Oncol J Eur Soc Ther Radiol Oncol*. 2009;93(2):372-376. doi:10.1016/j.radonc.2009.05.024

13. Dietzsch S, Braesigk A, Seidel C, et al. Pretreatment central quality control for craniospinal irradiation in non-metastatic medulloblastoma. *Strahlenther Onkol*. 2021;197(8):674-682. doi:10.1007/s00066-020-01707-8

14. Lin L, Dou Q, Jin YM, et al. Deep Learning for Automated Contouring of Primary Tumor Volumes by MRI for Nasopharyngeal Carcinoma. *Radiology*. 2019;291(3):677-686. doi:10.1148/radiol.2019182012

15. Wahid KA, Ahmed S, He R, et al. Evaluation of deep learning-based multiparametric MRI oropharyngeal primary tumor auto-segmentation and investigation of input channel effects: Results from a prospective imaging registry. *Clin Transl Radiat Oncol*. 2022;32:6-14. doi:10.1016/j.ctro.2021.10.003

16. Rodríguez Outeiral R, Bos P, Al-Mamgani A, Jasperse B, Simões R, van der Heide UA. Oropharyngeal primary tumor segmentation for radiotherapy planning on magnetic resonance imaging using deep learning. *Phys Imaging Radiat Oncol*. 2021;19:39-44. doi:10.1016/j.phro.2021.06.005

17. Breto AL, Spieler B, Zavala-Romero O, et al. Deep Learning for Per-Fraction Automatic Segmentation of Gross Tumor Volume (GTV) and Organs at Risk (OARs) in Adaptive Radiotherapy of Cervical Cancer. *Front Oncol*. 2022;12:854349. doi:10.3389/fonc.2022.854349

18. Cardenas CE, McCarroll RE, Court LE, et al. Deep Learning Algorithm for Auto-Delineation of High-Risk Oropharyngeal Clinical Target Volumes with Built-in Dice Similarity Coefficient Parameter Optimization Function. *Int J Radiat Oncol Biol Phys*. 2018;101(2):468. doi:10.1016/j.ijrobp.2018.01.114

19. Menze BH, Jakab A, Bauer S, et al. The Multimodal Brain Tumor Image Segmentation Benchmark (BRATS). *IEEE Trans Med Imaging*. 2015;34(10):1993-2024. doi:10.1109/TMI.2014.2377694

20. Zeng K, Bakas S, Sotiras A, et al. Segmentation of Gliomas in Pre-operative and Post-operative Multimodal Magnetic Resonance Imaging Volumes Based on a Hybrid Generative-Discriminative Framework. *Brainlesion Glioma Mult Scler Stroke Trauma Brain Inj BrainLes Workshop*. 2016;10154:184-194. doi:10.1007/978-3-319-55524-9_18

21. Ermiş E, Jungo A, Poel R, et al. Fully automated brain resection cavity delineation for radiation target volume definition in glioblastoma patients using deep learning. *Radiat Oncol Lond Engl*. 2020;15:100. doi:10.1186/s13014-020-01553-z

22. Bi N, Wang J, Zhang T, et al. Deep Learning Improved Clinical Target Volume Contouring Quality and Efficiency for Postoperative Radiation Therapy in Non-small Cell Lung Cancer. *Front Oncol*. 2019;9:1192. doi:10.3389/fonc.2019.01192

23. Frija G, Blažić I, Frush DP, et al. How to improve access to medical imaging in low- and middle-income countries ? *eClinicalMedicine*. 2021;38. doi:10.1016/j.eclinm.2021.101034

24. Ekpo EU, Egbe NO, Inyang SO, Azogor WE, Upeh ER. Challenges of radiological equipment management policies in some northern Nigerian hospitals. *South Afr Radiogr*. 2013;51(1):19-22.

25. Baskin JL, Lezcano E, Kim BS, et al. Management of children with brain tumors in Paraguay. *Neuro-Oncol*. 2013;15(2):235-241. doi:10.1093/neuonc/nos291

26. Taylor RE, Bailey CC, Robinson KJ, et al. Impact of radiotherapy parameters on outcome in the International Society of Paediatric Oncology/United Kingdom Children's Cancer Study Group PNET-3 study of preradiotherapy chemotherapy for M0-M1 medulloblastoma. *Int J Radiat Oncol Biol Phys*. 2004;58(4):1184-1193. doi:10.1016/j.ijrobp.2003.08.010

27. Pediatric Radiation Oncology - Louis S. Constine, Nancy J. Tarbell, Edward C. Halperin - Google Books. Accessed July 20, 2021. https://books.google.com/books?hl=en&lr=&id=j9ufDAAAQBAJ&oi=fnd&pg=PT24&ots=N5silhjyEl&sig=ZBUfI0bmT-CVHF0khXxAvoIVJwM#v=onepage&q&f=false

28. Isensee F, Jaeger PF, Kohl SAA, Petersen J, Maier-Hein KH. nnU-Net: a self-configuring method for deep learning-based biomedical image segmentation. *Nat Methods*. 2021;18(2):203-211. doi:10.1038/s41592-020-01008-z

29. Software in radiation therapy and oncology. RaySearch Laboratories. Accessed October 11, 2022. https://www.raysearchlabs.com/

30. RayStation product configurations. RaySearch Laboratories. Accessed October 11, 2022. https://www.raysearchlabs.com/products/raystation/safety-and-performance-information_raystation/

31. xxxxxxxxxxxxxxxxxxxxxxxxxxxxxxxxxxxxxxxxxxxxxxxxxxxxxxxxxxxxxxxxxxxxxxxxxxxxxx

32. Men K, Chen X, Zhang Y, et al. Deep Deconvolutional Neural Network for Target Segmentation of Nasopharyngeal Cancer in Planning Computed Tomography Images. *Front Oncol*. 2017;7. Accessed October 13, 2022. https://www.frontiersin.org/articles/10.3389/fonc.2017.00315

33. Mei H, Lei W, Gu R, et al. Automatic segmentation of gross target volume of nasopharynx cancer using ensemble of multiscale deep neural networks with spatial attention. *Neurocomputing*. 2021;438:211-222. doi:10.1016/j.neucom.2020.06.146

34. Savjani RR, Lauria M, Bose S, Deng J, Yuan Y, Andrearczyk V. Automated Tumor Segmentation in Radiotherapy. *Semin Radiat Oncol*. 2022;32(4):319-329. doi:10.1016/j.semradonc.2022.06.002

## FIGURES AND FIGURE LEGENDS

**FIGURE 1** Outline of the visibility metric calculation workflow. The image is cropped based on a 3D expansion of the clinically defined GTV mask. High-intensity pixel values are overwritten to NaN. Visibility metrics are calculated using a NaN mean.

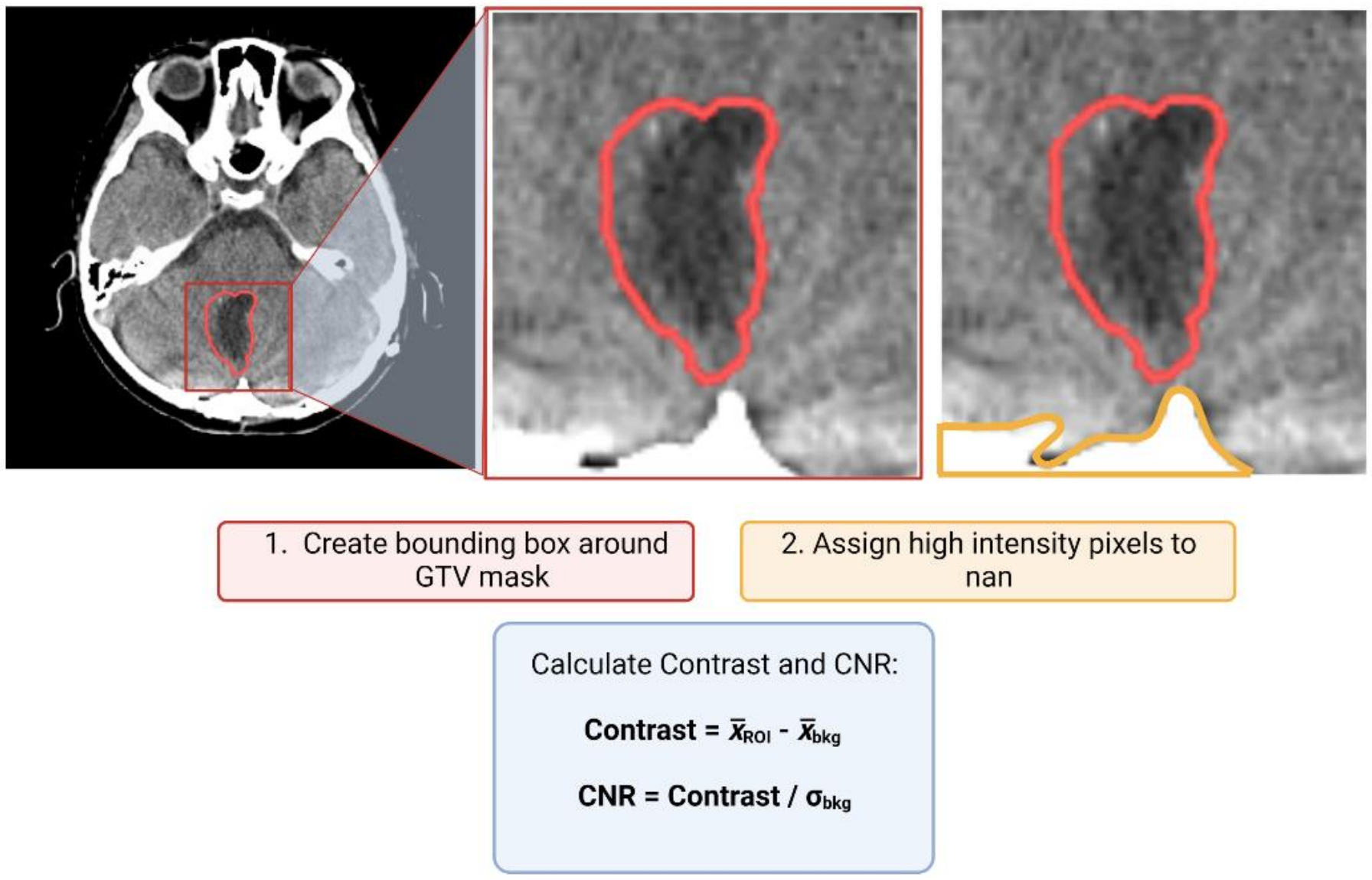

**Figure 2.** Overview of the four experiments compared in the visibility study. Model 1 was trained on the full dataset, and model 2 was trained on the dataset pre-selected for contour visibility. Both models were tested on two testing datasets that had or had not been pre-selected for contour visibility.

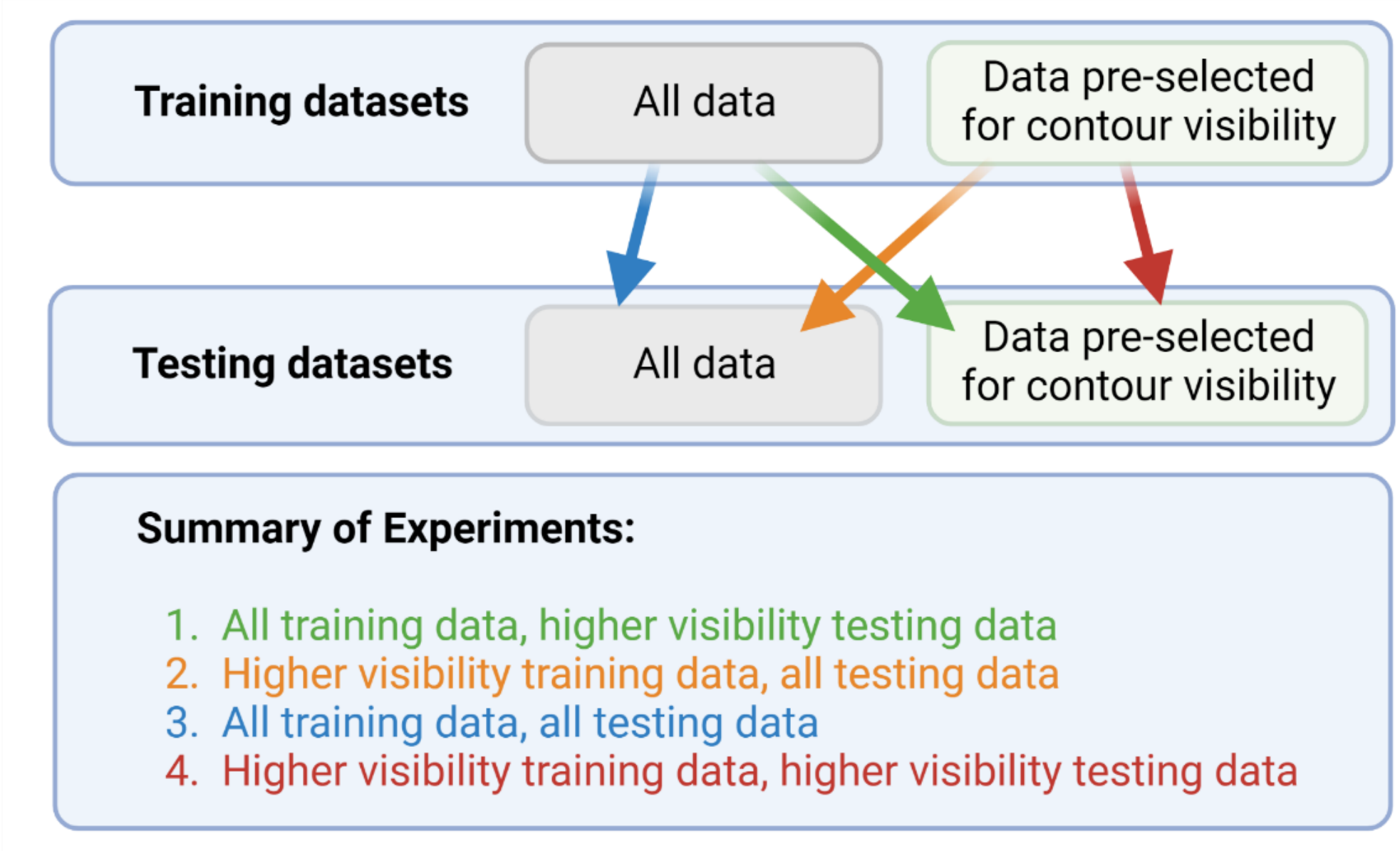

**FIGURE 3** Example results of the visual scoring system that was applied: 1 = <25% of resection cavity slices visible, 2 = 25%-50% of resection cavity slices clearly visible, 3 = 50%-75% of resection cavity slices clearly visible, and 4 = >75%-100% of resection cavity slices clearly visible.

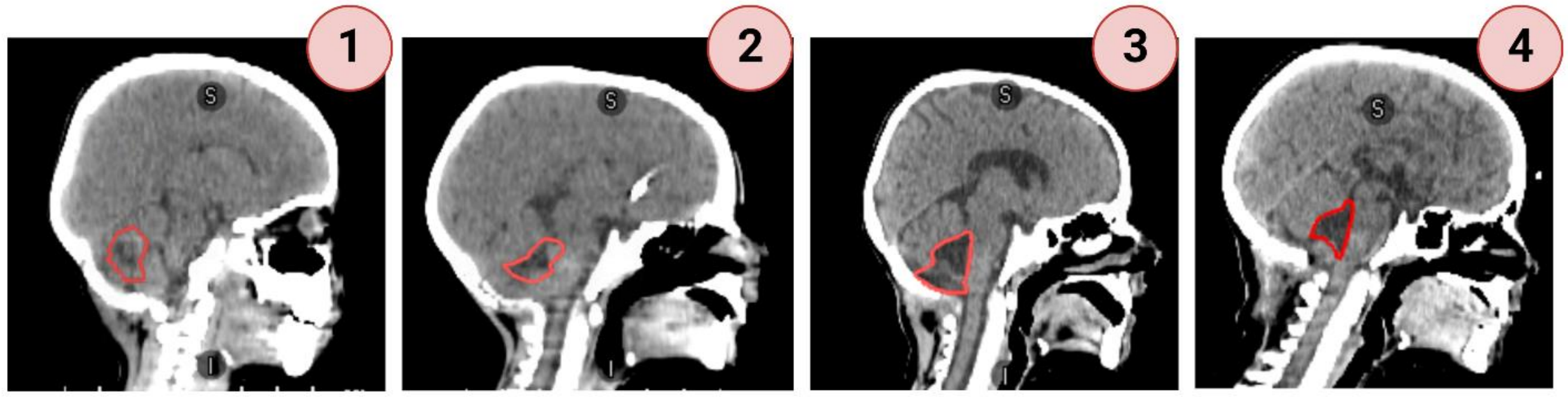

**FIGURE 4** Summary of qualitative and quantitative visibility metrics. (a) Calculated contrast values plotted against visibility scores. The red lines correspond to -10 HU and +10 HU, which were used as the thresholds for pre-selection. (b) CNR plotted against visibility scores. The red lines correspond to -0.5 and 0.1, which were used as the thresholds for CNR-based pre-selection. (c) Summary of the number of patients before removing low-visibility patients and after removing patients based on calculated contrast and CNR.

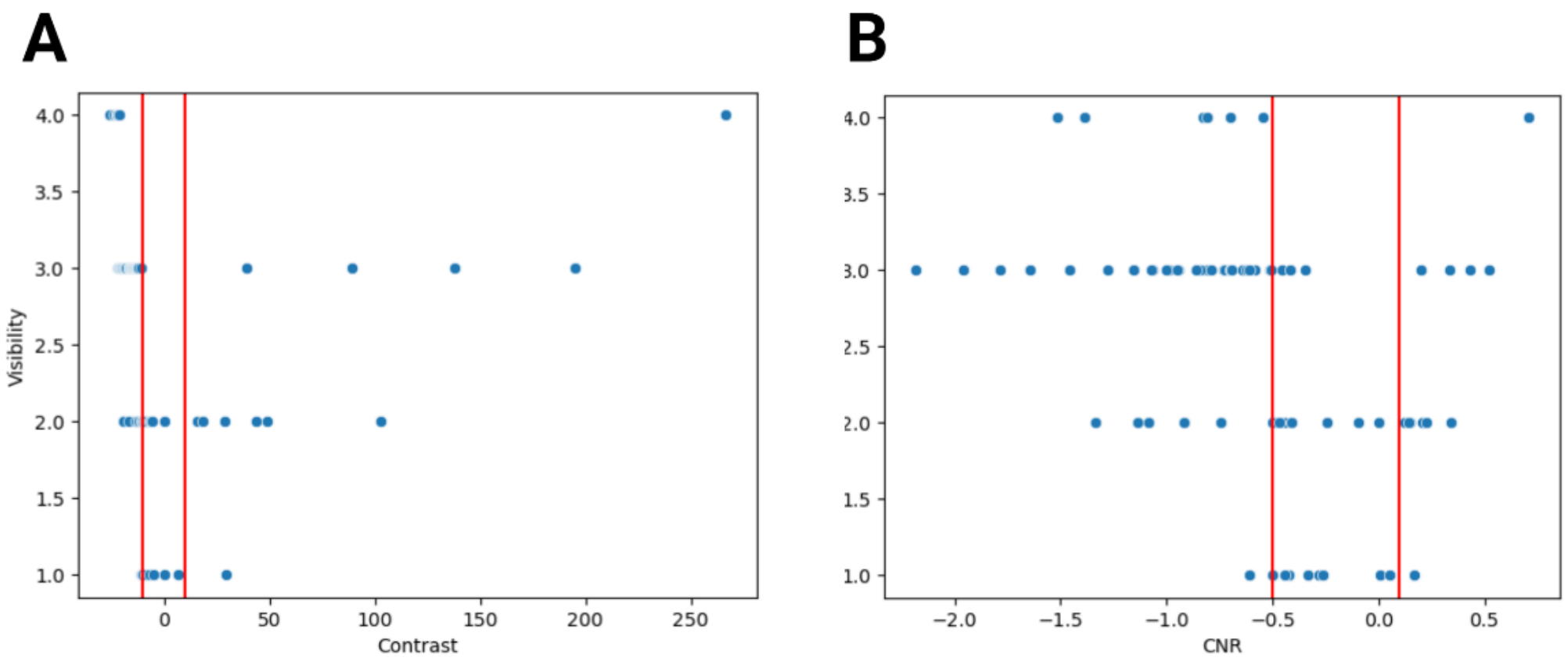

C

| Visibility Scores | No Filter | Contrast Filter | CNR Filter |
|---|---|---|---|
| 4 | 7 | 7 | 7 |
| 3 | 41 | 41 | 38 |
| 2 | 21 | 15 | 12 |
| 1 | 11 | 6 | 3 |
| Total Patients | 80 | 69 | 60 |

**FIGURE 5** Summary of surface and overlap metrics (DSC, HD [mm], and MSD [mm]) achieved by the GTV auto-contouring models tested on all data and higher visibility data. Two models were trained, one using all data and one using data pre-selected for visibility. Both models were then tested on two datasets, the full dataset and a dataset pre-selected for visibility.

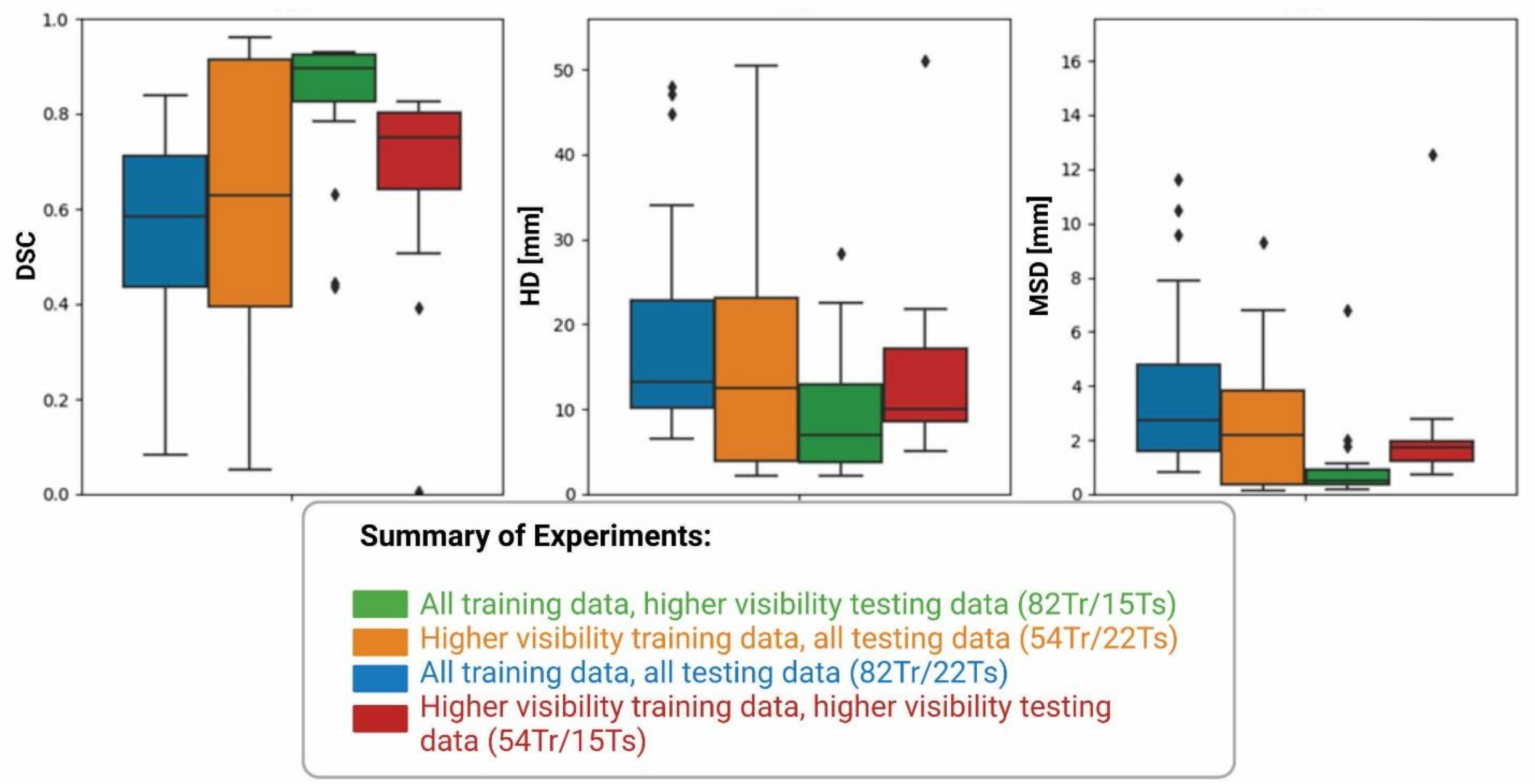

**FIGURE 6** (a) DSC, HD [mm], and MSD [mm] achieved in 15 test patients for the clinical vs. autocontoured GTV volumes (green) and the corresponding CTV volumes (purple). (b) Left, example of a GTV DSC of 0.62. After expansion, the DSC between the two corresponding CTV volumes improved to 0.82. Right, example of a GTV DSC of 0.92. After expansion, the DSC between the two corresponding CTV volumes improved to 0.98. The clinical GTV volume and its corresponding CTV (clinical GTV + 1.5 cm) are defined in dark green and dark purple, respectively. The autocontoured GTV and corresponding CTV (auto GTV + 1.5 cm) are defined in bright green and bright purple.

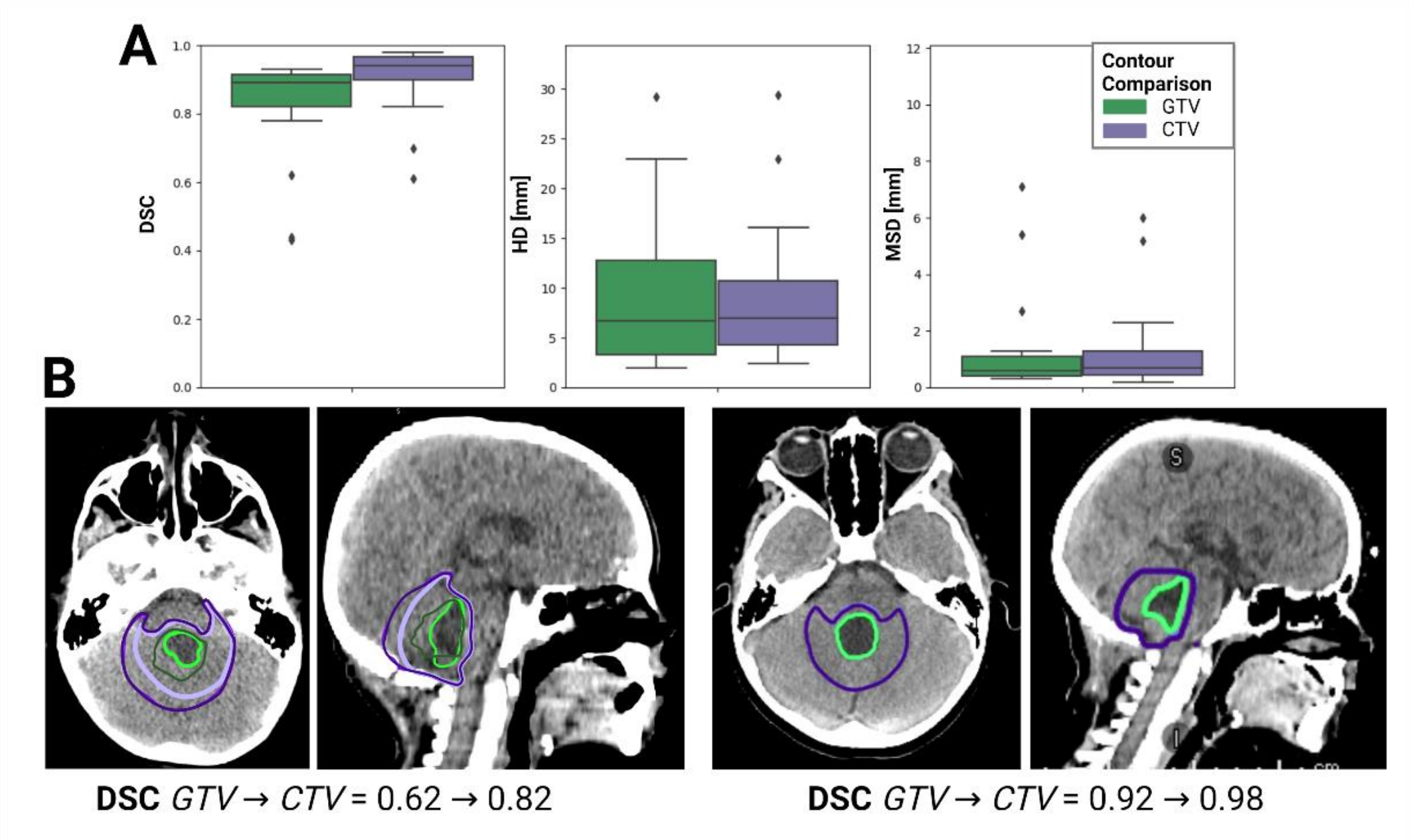